\documentclass[12pt]{article}
\begin{document}

\begin{center}
{\large \bf Existing and expected manifestations of a new fundamental interaction}
\vspace{0.5 cm}

\begin{small}
\renewcommand{\thefootnote}{*}
L.M.Slad\footnote{slad@theory.sinp.msu.ru} \\
{\it Skobeltsyn Institute of Nuclear Physics,
Lomonosov Moscow State University, Moscow 119991, Russia}
\end{small}
\end{center}

\vspace{0.3 cm}

\begin{footnotesize}
A number of characteristics of the new fundamental interaction are described. The interaction is carried by a massless pseudoscalar boson and extends to at least the electron neutrino, proton, and neutron. A substantiation of the existence of such an interaction is supported by an good agreement between the theoretical and experimental rates of all the five observed processes with solar neutrinos. A bright manifestation of the new interaction is expected in the observation that its contribution to the rate of splitting of a number of light stable nuclei by reactor antineutrinos is approximately six orders of magnitude greater than the contribution of electroweak interaction.
\end{footnotesize}

\vspace{0.5 cm}

\begin{small}

\begin{center}
{\large \bf 1. Introduction}
\end{center} 

A new fundamental interaction turned able to appear in a logically simple and clear form as a result of numerous works that have led on the basis of known interactions to the creation of the standard solar model (SSM), including the fluxes and the spectra of solar neutrinos from various sources. The creation of SSM was followed by theoretical calculations of the rates of a number of solar neutrino processes and by setting remarkable experiments that had completely refuted the theoretical expectations.

The results of the first experiment on solar neutrino registration by the 
transitions $\nu_{e}+{}^{37}{\rm Cl} \rightarrow e^{-}+{}^{37}{\rm Ar}$ \cite{1} have led the connoisseurs into confusion. Davis and colleagues stated the upper limit for such a transition rate as 3 SNU (1 SNU is $10^{-36}$ captures per target atom per second), while the rate predicted by Bahcall \cite{2} was $30^{+30}_{-15}$ SNU.

Soon, the emergence of the solar neutrino problem was assigned to the existence of neutrino oscillations in vacuum, a hypothesis of which was proposed by Gribov and Pontecorvo \cite{3}. This opinion, transforming, turned over time into an unshakable faith. Such a transformation, attributed to neutrinos in the medium, was basically formulated by Wolfenstein \cite{4}, and subsequently received wide recognition thanks to the Mikheev-Smirnov work \cite{5}. It includes, in addition to doubling the initial number of neutrinos necessary for introducing oscillations, the assumption of a coherent nature of the elastic forward scattering of solar neutrinos by the electrons of the medium. 

The results of the observed processes with solar neutrinos in their integrity were a peculiar Procrustean bed for the neutrino oscillation hypothesis. So, as it was shown in the work \cite{6}, the Mikheev-Smirnov-Wolfenstein mechanism with parameters from SNO \cite{7} and Super-Kamiokande \cite{8}, taking into account the volume distribution of neutrino sources in the Sun \cite{9}, contradicts the results of four out of five observed processes with solar neutrinos. 

The first work giving an alternative solution to the solar neutrino problem and based on logically clear methods of the classical field theory has been publicly presented in February 2015 \cite{10}. In this work, the existence of a new interaction mediated by a massless pseudoscalar boson with Yukawa coupling to electron neutrinos and nucleons is postulated. With only a single free parameter, a good agreement is demonstrated between the theoretical and experimental results for four different processes with solar neutrinos. The probability that this agreement is a play of chance rather than the true nature of things related to neutrinos seems negligible. This was subsequently confirmed \cite{11} by a good agreement between the theoretical and experimental numbers concerning the process of deuteron disintegration by neutral currents of solar neutrinos, which remained not considered in the work \cite{10}. The first publication on the solar neutrino problem as evidence of the existence of a new interaction, which includes all the main provisions and results of \cite{10} and \cite{11}, has appeared only in December 2019 \cite{12}.

\begin{center}
{\large \bf 2. Some aspects of the new interaction}
\end{center}
 
The new fundamental interaction, hereafter named semi-weak, is the Yukawa interaction of a hypothetical massless pseudoscalar boson with at least electron neutrinos and nucleons, but not with electrons (at the tree level), and is described by the following Lagrangian
\begin{equation}
{\cal L} = ig_{\nu_{e}ps}\bar{\nu}_{e}\gamma^{5}\nu_{e}\varphi_{ps}+
ig_{Nps}\bar{p}\gamma^{5}p\varphi_{ps}-ig_{Nps}\bar{n}\gamma^{5}n\varphi_{ps}.
\label{1}
\end{equation}
The interaction of the pseudoscalar boson $\varphi_{ps}$ with neutrinos of different types may be the same or different.

We believe that an equivalent group-theoretical, relativistic-covariant description should be given to the charged leptons and their neutrinos. So, we assume that the field of an any type neutrino transforms as a bispinor representation of the proper Lorentz group and obeys the (almost) massless Dirac equation. All of the positive energy solutions of such an equation, including the left-handed and right-handed ones, refer to different states of the same neutrino. The Lorentzian equality between the fields of the electron and the electron neutrino is a necessary condition for constructing an initially P-invariant (logically corrected left-right symmetric) model of electroweak interaction, performed in the work \cite{13}. In this model, the physical vacuum does not have a certain P-parity, and the fields of all intermediate bosons are a superposition of polar and axial 4-vectors, and these vectors have the same weight in the fields {\boldmath $W$}-bosons. These remarkable P-properties of vacuum and fields could not be revealed in the standard left-right symmetric model \cite{14}, since no P-transforms were considered there.

Note now that the pseudoscalar current connects the left and right components of the Dirac bispinor: $\bar{\psi}(p_{2})\gamma^{5}\psi(p_{1})=\bar{\psi}_{L}(p_{2})\psi_{R}(p_{1})-\bar{\psi}_{R}(p_{2})\psi_{L}(p_{1})$. Because of this, when a neutrino with energy $\omega_{1}$ elastically scatters off a rest nucleon of mass $M$, as a result of the interaction (\ref{1}), its handedness changes (every time) from left to right or vice versa. The total cross section of this process is
\begin{equation}
\sigma = \frac{(g_{\nu_{e}ps}g_{Nps})^{2}}{16 \pi M^{2}}
\cdot \frac{1}{(1+2 \omega_{1}/M)}.
\label{2}
\end{equation}
For solar neutrinos with the energies less than approximately 18.8 MeV, the cross section (\ref{2}) can be, with sufficient accuracy, considered independent of the energy $\omega_{1}$. The energy of the scattered neutrino $\omega_{2}$ is then evenly distributed in the interval
\begin{equation}
\frac{\omega_{1}}{1+2\omega_{1}/M} \leq \omega_{2} \leq \omega_{1},
\label{3}
\end{equation}
i.e. the relative fraction of the average energy loss of the neutrino in a collision with the nucleon is proportional to its initial energy
\begin{equation}
\frac{\omega_{1}-\omega_{2}}{2\omega_{1}} = \frac{\omega_{1}}{M}\cdot \frac{1}{1+2\omega_{1}/M}.
\label{4}
\end{equation}

\begin{center}
{\large \bf 3. Qualitative consequences of the semi-weak interaction of solar neutrinos with nucleons}
\end{center} 

Semi-weak interaction manifests itself in the results of four out of five experiments with solar neutrinos, ${}^{37}{\rm Cl} \rightarrow {}^{37}{\rm Ar}$, ${}^{71}{\rm Ga} \rightarrow {}^{71}{\rm Ge}$, $\nu_{e} e^{-}\rightarrow \nu_{e} e^{-}$, and $\nu_{e}D \rightarrow  e^{-}pp$, only through the corollaries of the neutrino collisions with the nucleons of the Sun. There are two such corollaries. One of them consists in converting part of the initially left-handed neutrinos into right-handed ones, so that the effective flux of solar neutrinos reduces to the flux of left-handed neutrinos, since the contributions from right-handed neutrinos to the rates of the above processes are negligible in the initially P-invariant model \cite{13} and in the left-right symmetric model \cite{14} of electroweak interaction, and they are absent in the Weinberg-Salam model. The second consequence is the decrease in the energy of every neutrino compared to that gained at the production, as a result, the rates of the listed processes decrease, since the cross sections decrease with decreasing energy, each to its own law.

Among the above-mentioned experiments, a special position is taken by the gallium experiment from a theoretical point of view. In the transition from gallium to germanium, the dominant contribution is given by the neutrinos from $p-p$-collisions and from ${}^{7}{\rm Be}$, with the initial energy not exceeding 0.862 MeV. By virtue of relation (\ref{4}), the neutrino energy changes very little after about ten neutrino-nucleon collisions, and the difference between the results of SSM and of the experiment is determined, almost completely, by the relation between the fluxes of the left-handed and right-handed neutrinos. Since the experimental ${}^{71}{\rm Ga} \rightarrow {}^{71}{\rm Ge}$ transition rates are slightly smaller than half of the rate expected from SSM, then this indicates that the fluxes of left-handed and right-handed neutrinos at the Earth's surface are approximately equal.

The results of the fifth solar neutrino experiment, $\nu_{e}D \rightarrow \nu_{e} np$, reflect the manifestation of semi-weak interaction of the electron neutrino not only in the Sun, but also in the terrestrial installation. Namely, the disintegration of the deuteron into a proton and a neutron is due to two non-interfering sub-processes. The first sub-process, which has a standard description based on the Weinberg-Salam model, involves only left-handed solar neutrinos, which interact with the deuteron nucleons due to $Z$-boson exchange. In the second subprocess, induced by the semi-weak interaction, both left-handed and right-handed solar neutrinos are involved.

An enough precise description of the solar neutrino spectrum at the Earth's surface, with the collisions of neutrinos with the nucleons of the Sun due to semi-weak interaction (\ref{1}) being taken into account, is an extremely difficult task, which consists in numerical calculations of the relevant consequences of the Brownian motion of the neutrino from its production moment till the exit from the Sun. Such calculations should take into account the distributions of nucleons and neutrino sources over the distance from the center of the Sun obtained in the framework of SSM \cite{9}. The only free parameter of the theory regulating these consequences is the product of the coupling constants in Lagrangian (\ref{1}), $\beta \equiv g_{\nu_{e}ps}g_{Nps}/4\pi$. The analyzing procedure should consist in obtaining the distribution $P_{\beta}(s,n)$ in the number of neutrino-nucleon collisions inside the Sun $n$ for every $\beta$ (from a set given in advance), with $P_{\beta}(s,n)$ depending on the neutrino source $s$. If we limit the number of collisions to about ten,then by virtue of the ratio (\ref{2}) the distribution $P_{\beta}(s,n)$ can be in the first approximation considered independent of the initial neutrino energy. Combining the distribution $P_{\beta}(s,n)$ with the kinematics of a single collision of the electron neutrino with the nucleon, as described by equation (\ref{3}), could yield the spectrum at the Earth's surface for the left- and right-handed solar neutrinos produced by the source $s$ with initial energy $\omega_{1}$. The chain of all theoretical calculations would have to end up with finding the constant $\beta$ best fitting the results of all solar neutrino experiments.

The exact value of the rate $V_{A}$ of that or another observed process $A$, caused by the Weinberg-Salam electroweak interaction of left-handed solar neutrinos from the source $s$ with the flux $\Phi_{s}$, is expressed through the distribution $P_{\beta}(s, 2k)$ and partial rates $v_{A}(2k)$, originating from neutrinos that underwent $2k$ collisions with nucleons of the Sun, by a relation of the form
\begin{equation}
V_{A} = \Phi_{s} \sum_{k=0}^{+\infty} P_{\beta}(s, 2k) v_{A}(2k).
\label{5}
\end{equation}
Unable to accurately calculate the distribution of $P_{\beta}(s, n)$, we make in the formula (\ref{5}) the following logically acceptable replacement of the its right-hand side
\begin{equation}
V_{A} = \Phi_{s} [c v_{A}(2k_{1})+ (2-c) v_{A}(2k_{1}+2)] \sum_{k=0}^{+\infty} P_{\beta}(s, 2k),
\label{6}
\end{equation}
where $0 \leq c \leq 2$. According to the assumption made in \cite{12}, the probabilities of an even and an odd number of collisions of neutrinos with nucleons of the Sun are approximately equal, i.e. they have a value of 0.5. The quantities $ k_{1}$ and $c$ in the formula (\ref{6}) are two free parameters.

Let us introduce the partial rates $v_{A}(2k + 1)$, which could be generated by left-handed neutrinos from a source $s$ with an energy spectrum identical to the energy spectrum of right-handed neutrinos that appeared after $2k + 1$ collisions with nucleons of the Sun. We find it admissible to take the following approximation
\begin{equation}
v_{A}(2k+1) = \frac{1}{2}(v_{A}(2k) + v_{A}(2k+2)).
\label{7}
\end{equation}
Using the above, the relation (\ref{6}) can be written as
\begin{equation}
V_{A} = 0.5 \Phi_{s} \cdot \left\{
\begin{array}{ll}
2 v_{A}(2k_{1}+1)+ (1-c) v_{A}(2k_{1}+2), & {\mbox{\rm если}} \hspace{0.3cm}
c \in [0,1], \\
(c-1) v_{A}(2k_{1})+ 2 v_{A}(2k_{1}+1), & {\mbox{\rm если}} \hspace{0.3cm}
c \in [1,2].
\end{array} \right.
\label{8}
\end{equation}
We are able to give two parts of the formula (\ref{8}) in a single simple form
\begin{equation}
V_{A} = 0.5 \Phi_{s} [(1-a) v_{A}(n_{0}) + a v_{A}(n_{0}+1)],
\label{9}
\end{equation}
where $0 \leq a < 1$. We assume that the integer, even or odd, number $n_{0}$ and number $a$ are the same for all solar neutrino sources $s$ and for all observed processes $A$ and energy constraints in experiments, if they exist. Posterior assessment of this assumption is given by good agreement between the theoretical and experimental characteristics of the observed processes. The value $V_{A}$ is defined in the relation (\ref{9}) in fact by one parameter, $n_{a} \equiv n_{0} + a$, which we call the effective collision number and which can be either integer or non-integer.

For the deuteron disintegration, caused by the semi-weak interaction (\ref{1}) of both left- and right-handed solar neutrinos, the relation (\ref{9}) with the omitted factor 0.5 is obviously also valid.

In the work \cite{12}, where the central values of the BP04 model \cite {15} are taken for neutrino fluxes from all sources, including ${}^{8}{\rm B}$, we show that a good agreement between the theoretical and all experimental results is achieved at $n_{0} = 11$, $a = 0$. This made it possible to obtain an estimate of the product of the coupling constants, expressed by the value
\begin{equation}
\beta \equiv g_{\nu_{e}ps}g_{Nps} /4\pi = (3.2 \pm 0.2)\times 10^{-5}.
\label{10}
\end{equation}

This paper presents the results of calculating the rates of processes with solar neutrinos at the effective number of collisions $n_{a} = 11.5$ and at the neutrino flux $\Phi$ from ${}^{8}{\rm B}$ equal to $6.01 \times 10^{6}(1 \pm 0.23)$ ${\rm cm}^{-2}{\rm s}^{-1}$.

\begin{center}
{\large \bf 4. Quantitative consequences of the semi-weak interaction of solar neutrinos with nucleons}
\end{center} 

The procedure of calculating the rates of all the five observed processes, described in detail in the work \cite{12}, is fully preserved in our present consideration.

The process which has started the solar neutrino problem is $\nu_{e}+{}^{37}{\rm Cl} \rightarrow e^{-}+{}^{37}{\rm Ar}$. The threshold energy is 0.814 MeV. The experimental value of the process rate is $2.56 \pm 0.16 \pm 0.16$ SNU \cite{16}.
\begin{center}
{\bf Table 1.} The rates of transitions ${}^{37}{\rm Cl} \rightarrow {}^{37}{\rm Ar}$ in SNU.
\begin{tabular}{lccccccc}
\multicolumn{8}{c}{The flux $\Phi$ is in units of $10^{6}$ ${\rm cm}^{-2}{\rm s}^{-1}$} \\ 
\hline
\multicolumn{1}{l}{} 
&\multicolumn{1}{c}{${}^{8}{\rm B}$} 
&\multicolumn{1}{c}{${}^{7}{\rm Be}$}
&\multicolumn{1}{c}{${}^{15}{\rm O}$}
&\multicolumn{1}{c}{$pep$}
&\multicolumn{1}{c}{${}^{13}{\rm N}$}
&\multicolumn{1}{c}{$hep$}
&\multicolumn{1}{c}{Total} \\ 
\hline
SSM \cite{12} & 6.1 & 1.1 & 0.3 & 0.2 & 0.1 & 0.03 & 7.9 \\
Eq. (\ref{1}), $n_{0} = 11$, $\Phi =5.79$  & 1.97 & 0.43 & 0.17 & 0.11 & 0.04 & 0.01 & 2.72 \\
Eq. (\ref{1}), $n_{a} = 11.5$, $\Phi =6.01$  & 2.00 & 0.42 & 0.17 & 0.11 & 0.04 & 0.01 & 2.75 \\
\hline
\end{tabular}
\end{center}
 
The process is $\nu_{e}+{}^{71}{\rm Ga} \rightarrow e^{-}+{}^{71}{\rm Ge}$. The threshold energy is 0.233 MeV. The experimental values of the process rates are $62.9^{+6.0}_{-5.9}$ SNU \cite{17} and $65.4^{+3.1}_{-3.0}{}^{+2.6}_{-2.8}$ SNU \cite{18}.
\begin{center}
{\bf  Table 2.} The rates of transitions ${}^{71}{\rm Ga} \rightarrow {}^{71}{\rm Ge}$ in SNU.
\begin{tabular}{lcccccccc}
\multicolumn{8}{c}{The flux $\Phi$ is in units of $10^{6}$ ${\rm cm}^{-2}{\rm s}^{-1}$} \\ 
\hline
\multicolumn{1}{l}{}
&\multicolumn{1}{c}{$p$-$p$}
&\multicolumn{1}{c}{${}^{7}{\rm Be}$} 
&\multicolumn{1}{c}{${}^{8}{\rm B}$} 
&\multicolumn{1}{c}{${}^{15}{\rm O}$}
&\multicolumn{1}{c}{${}^{13}{\rm N}$}
&\multicolumn{1}{c}{$pep$}
&\multicolumn{1}{c}{$hep$}
&\multicolumn{1}{c}{Total} \\ 
\hline
SSM \cite{9} & 70.8 & 34.3 & 14.0 & 6.1 & 3.8 & 3.0 & 0.06 & 132 \\
Eq. (\ref{1}), $n_{0} = 11$, $\Phi =5.79$ & 34.6 & 17.2 & 4.9 & 2.8 & 1.7 & 1.4 & 0.02 & 62.6 \\
Eq. (\ref{1}), $n_{a} = 11.5$, $\Phi =6.01$ & 34.6 & 17.1 & 5.0 & 2.8 & 1.7 & 1.4 & 0.02 & 62.6 \\
\hline
\end{tabular}
\end{center}

The process of elastic scattering of solar neutrinos on electrons was studied in Super-Kamio- kande (SK) \cite{19}--\cite{21} and in Sudbury Neutrino Observatory (SNO) \cite{22}--\cite{24}, \cite{7}. The lower limit for the reconstructed energy of the scattered electron $E_{c}$ is set from experimental considerations.
\begin{center}
{{\bf Table 3.} Effective fluxes of neutrinos $\Phi_{eff}^{\nu e}$ found from the process $\nu_{e} e^{-}\rightarrow \nu_{e} e^{-}$} \\
\begin{tabular}{lcccc}
\multicolumn{5}{c}{($E_{c}$ are given in MeV and the fluxes are in units of $10^{6}$ ${\rm cm}^{-2}{\rm s}^{-1}$).} \\ 
\hline
\multicolumn{1}{l}{References}
&\multicolumn{1}{c}{$E_{c}$}  
&\multicolumn{1}{c}{Experi-}
&\multicolumn{1}{c}{Eq. (\ref{1}),}
&\multicolumn{1}{c}{Eq. (\ref{1}),} \\
& & mental & $n_{0} = 11$, $\Phi =5.79$ & $n_{a} = 11.5$, $\Phi =6.01$ \\
& & $\Phi_{eff}^{\nu e}$ & $\Phi_{eff}^{\nu e}$ & $\Phi_{eff}^{\nu e}$
\\
\hline
SK III \cite{21} & 5.0 &$2.32\pm 0.04\pm 0.05$ & 2.27 & 2.33  \\
SK II \cite{20} & 7.0 &$2.38\pm 0.05{}^{+0.16}_{-0.15}$ & 2.00 & 2.04 \\
SK I \cite{19} & 5.0 &$2.35\pm 0.02\pm 0.08$ & 2.27 & 2.33 \\
SNO III \cite{7} & 6.5 &$1.77^{+0.24}_{-0.21}{}^{+0.09}_{-0.10}$ & 2.01 & 2.05 \\
SNO IIB \cite{24} & 6.0 &$2.35\pm 0.22\pm 0.15$ & 2.10 & 2.15 \\
SNO IIA \cite{23} & 6.0 &$2.21^{+0.31}_{-0.26}\pm 0.10$ & 2.10 & 2.15 \\
SNO I \cite{22} & 5.5 &$2.39^{+0.24}_{-0.23}{}^{+0.12}_{-0.12}$ & 2.19 & 2.24 \\
\hline
\end{tabular}
\end{center}

The process of deuteron disintegration by the charged solar neutrino current is $\nu_{e}+D \rightarrow e^{-}+p+p$. At the different phases of the experiment in SNO, the reconstructed energy of the produced electron was limited from below by different values of $E_{c}$.
\begin{center}
{{\bf Table 4.} Effective fluxes of neutrinos $\Phi_{eff}^{cc}$ found from the process $\nu_{e}D \rightarrow  e^{-}pp$} \\
\begin{tabular}{lcccc}
\multicolumn{5}{c}{($E_{c}$ are given in MeV and the fluxes are in units of $10^{6}$ ${\rm cm}^{-2}{\rm s}^{-1}$).} \\ 
\hline
\multicolumn{1}{l}{References}
&\multicolumn{1}{c}{$E_{c}$}  
&\multicolumn{1}{c}{Experi-}
&\multicolumn{1}{c}{Eq. (\ref{1}),}
&\multicolumn{1}{c}{Eq. (\ref{1}),} \\
& & mental & $n_{0} = 11$, $\Phi =5.79$ & $n_{a} = 11.5$, $\Phi =6.01$ \\
& & $\Phi_{eff}^{cc}$ & $\Phi_{eff}^{cc}$ & $\Phi_{eff}^{cc}$
\\
\hline
SNO III \cite{7} & 6.5 & $1.67^{+0.05}_{-0.04}{}^{+0.07}_{-0.08}$ & 1.67 & 1.67 \\
SNO IIB \cite{24} & 6.0 & $1.68^{+0.06}_{-0.06}{}^{+0.08}_{-0.09}$ & 1.78 & 1.79 \\
SNO IIA \cite{23} & 6.0 & $1.59^{+0.08}_{-0.07}{}^{+0.06}_{-0.08}$ & 1.78 & 1.79 \\
SNO I \cite{22} & 5.5 & $1.76^{+0.06}_{-0.05}{}^{+0.09}_{-0.09}$ & 1.88 & 1.90 \\
\hline
\end{tabular}
\end{center}

The process of deuteron disintegration by neutral solar neutrino currents is $\nu_{e}+D \rightarrow \nu_{e}+n+p$. The experimental effective neutrino flux values $\Phi_{eff}^{nc}$ found from this process (in units of $10^{6}$ ${\rm cm}^{-2}{\rm s}^{-1}$) are: 
$$5.09^{+0.44}_{-0.43}{}^{+0.46}_{-0.43} \; \cite{22}, \quad 5.21 \pm 0.27 \pm 0.38 \; \cite{23}, \quad 4.94^{+0.21}_{-0.21}{}^{+0.38}_{-0.34} \; \cite{24}, \quad 5.54^{+0.33}_{-0.31}{}^{+0.36}_{-0.34} \; \cite{7}.$$
The theoretical value of the effective neutrino flux $\Phi_{eff}^{nc}$ for the considered process is the sum of the effective neutrino fluxes $\Phi_{eff}^{Z}$ and $\Phi_{eff}^{\varphi_{ps}}$, which correspond to two non-interfering sub-processes of the disintegration of the deuteron into a neutron and a proton due to the exchange of the boson $Z$ and of the pseudoscalar boson $\varphi_{ps}$, respectively.
\begin{center}
{{\bf Table 5.} The theoretical effective fluxes $\Phi_{eff}^{Z}$, $\Phi_{eff}^{\varphi_{ps}}$, and $\Phi_{eff}^{nc}$}\\
\begin{tabular}{cccc}
\multicolumn{4}{c}{(in units of $10^{6}$ ${\rm cm}^{-2}{\rm s}^{-1}$).} \\ 
\hline
\multicolumn{1}{c}{}
&\multicolumn{1}{c}{$\Phi_{eff}^{Z}$}
&\multicolumn{1}{c}{$\Phi_{eff}^{\varphi_{ps}}$} 
&\multicolumn{1}{c}{$\Phi_{eff}^{nc}$} \\
\hline
Eq. (\ref{1}), $n_{0} = 11$, $\Phi =5.79$ & 2.10 & $2.87 \pm 0.36$ & $4.97 \pm 0.36$ 
\\
Eq. (\ref{1}), $n_{a} = 11.5$, $\Phi =6.01$ & 2.15 & $2.96 \pm 0.37$ & $5.11 \pm 0.37$ \\
\hline
\end{tabular}
\end{center}

The most remarkable and expected consequence of coming from the effective number of collisions $n_{0} = 11$ at the neutrino flux from ${}^{8}{\rm B}$ equal to 5.79 ${\rm cm}^{-2}{\rm s}^{-1}$ to the effective number of collisions $n_{a} = 11.5$ at the flux equal to 6.01 ${\rm cm}^{-2} {\rm s}^{-1}$ is the increase in the theoretical effective fluxes for the $\nu_{e} e^{-}\rightarrow \nu_{e} e^{-}$ process with almost unchanged fluxes for the $\nu_{e}D \rightarrow e^{-}pp$ process.

\begin{center}
{\large \bf 5. On the splitting of a number of stable nuclei by reactor antineutrinos caused by the semi-weak interaction}
\end{center}
 
As it is noted above, non-interfering contributions of electroweak and semi-weak interactions to the rate of deuteron disintegration by neutral currents of solar neutrinos are comparable. We draw attention to the fact that the Yukawa coupling constants of pseudoscalar boson $\varphi_{ps}$ to the proton and the neutron in the Lagrangian (\ref{1}) are opposite, so that the cross section of the deuteron disintegration sub-process with $\varphi_{ps}$ exchange has a decreasing factor $((M_{n}-M_{p})/M)^{2}$.

The presence of two comparable contributions to the deuteron disintegration rate by neutral currents of antineutrino reactors could be tested in an experiment located near the reactor. In case of a positive outcome of such an experiment, the hypothesis of semi-weak interaction would be out of competition.

A detailed discussion of the qualitative and quantitative aspects of the deuteron disintegration by reactor antineutrinos, as well as the problem of neutron registration is given in Ref. \cite{25} with the completed reactor experiments \cite{26}, \cite{27} taken as an example in comparing their methods with the methods \cite{28}, \cite{29} of experiments with solar neutrinos.

Note, that practically irrespective of the number of fissions of the different isotopes in the reactor, the contribution from the sub-process with the pseudoscalar boson $\varphi_{ps}$ exchange to the deuteron disintegration rate is about three times greater than that from the sub-process with the exchange by $Z$ boson exchange, i.e., we expect the observed desintegration rate to be approximately 4 times higher than the rate calculated in the electroweak interaction models. At the same time, the expected rate of the $\bar{\nu}_{e}D \rightarrow \bar{\nu}_{e}np$ process should be approximately 11--12 times higher than the $\bar{\nu}_{e}D \rightarrow e^{+}nn$ process rate.

A special impression is generated by the relations between the contributions from the semi-weak and electroweak interactions to the splitting rate of a number of stable nuclei, $^{3}$He, $^{7}$Li, $^{9}$Be and $^{19}$F, induced by neutral currents of the reactor antineutrinos. As a result of unequal numbers of the protons and neutrons in the above nuclei, there will not be the decreasing factor $((M_{n}-M_{p})/M)^{2}$ in the cross section for the nucleus splitting sub-process caused by the semi-weak interaction. As this factor made the two contributions to the deuteron disintegration rate comparable, its absence will make the $\varphi_{ps}$-boson exchange contribution to the rate of the nucleus splitting process dominating over the $Z$-boson exchange contribution by approximately six orders of magnitude.

To estimate the order of magnitude of the expected splitting rate of a nucleus ${\rm Y}$ containing $N_{\rm Y}$ neutrons and $Z_{\rm Y}$ protons by reactor antineutrinos, we take the $Z$-boson  exchange cross section, $\sigma_{\rm Z}^{\rm Y}$, and the $\varphi_{ps}$-boson exchange cross section, $\sigma_{ps}^{\rm Y}$, as  approximated through the respective deuteron disintegration cross sections, $\sigma_{\rm Z}^{\rm D}$ and $\sigma_{\rm ps}^{\rm D}$:
\begin{equation}
\sigma_{\rm Z}^{\rm Y}(\omega) = N_{\rm Y}^{2}\sigma_{\rm Z}^{\rm D}(\omega-E_{\rm th}^{\rm Y}+B),
\label{11}
\end{equation}
\begin{equation}
\sigma_{\rm ps}^{\rm Y}(\omega) = \left(\frac{M}{M_{n}-M_{p}}\right)^{2} 
(Z_{\rm Y}-N_{\rm Y})^{2}\sigma_{\rm ps}^{\rm D}(\omega-E_{\rm th}^{\rm Y}+B),
\label{12}
\end{equation}
where $B$ is the deuterium binding energy and $E_{\rm th}^{\rm Y}$ is the ${\rm Y}$-nucleus splitting threshold energy. These formulas, partially reflecting the coherence phenomenon, are very rough approximations. However, we hope that they reflect more or less correctly the threshold behavior of cross-sections and at least the order of magnitude of the cross-sections of the nucleus splitting. The cross sections $\sigma_{\rm Z}^{\rm D}(\omega)$ and $\sigma_{\rm ps}^{\rm D}(\omega)$ are taken from Refs. \cite{30} and \cite{25}, respectively.

The chain of the calculations concerning the estimates of the splitting rates for a number of stable nuclei is carried out according to the scheme described in Ref. \cite{25}. To find the splitting threshold energy for this or that nucleus, the values of the binding energies from the comprehensive work \cite{31} are used. For the relative contributions of the isotopes $^{235}{\rm U}$, $^{238}{\rm U}$, $^{239}{\rm Pu}$, and $^{241}{\rm Pu}$ to the number of fissions in a reactor, we accept their average values over the standard operation period of the VVER-1000 reactor \cite{32}: 0.56:0.07:0.31:0.06. We take the reactor antineutrino spectra from Ref. \cite{33}. The cross sections (\ref{11}) and (\ref{12}) of the nucleus splitting, integrated over the spectrum of reactor antineutrinos and averaged over the relative contributions of the reactor isotopes will be referred to as the weighted integral cross sections and denoted as $\Sigma^{\rm Y}_{\rm X}$, where ${\rm X = Z, ps}$. The contributions of the sub-processes to the nucleus splitting rate ${\cal N}^{\rm Y}_{\rm X}$ per mole of the respective atoms in a installation located at a distance $R$ from a reactor with the thermal power $W$ are found by the following formula
\begin{equation}
{\cal N}^{\rm Y}_{\rm X} = 1.58\cdot 10^{45} \cdot \Sigma^{\rm Y}_{\rm X} \cdot 
\frac{W}{\rm megawatt}\cdot \frac{1}{4\pi R^{2}} \cdot \frac{\rm fission}{{\rm mole}\cdot {\rm day}}.
\label{13}
\end{equation}
We give numerical estimates assuming $W = 2000$ megawatt and $R = 20$ m.

Taking into account the fact \cite{33} that the reactor antineutrino spectra, rapidly decreasing with energy, practically end at 8 MeV, we limit our analysis to the four light nuclei with unequal numbers of protons and neutrons whose splitting energy thresholds are lower than 6 MeV.

The first process is
\begin{equation}
\bar{\nu}_{e}+^{3}{\rm He} \rightarrow\bar{\nu}_{e}+{\rm D}+{\rm H}. 
\label{14}
\end{equation}
The binding energies of the nuclei $^{3}{\rm He}$ and D are 7.718 MeV and 2.225 MeV, respectively.

The second process is
\begin{equation}
\bar{\nu}_{e}+^{7}{\rm Li} \rightarrow\bar{\nu}_{e}+^{4}{\rm He}+^{3}{\rm H}. 
\label{15}
\end{equation}
The binding energies of the nuclei $^{7}{\rm Li}$, $^{4}{\rm He}$, and $^{3}{\rm H}$ are 39.245 MeV, 28.296 MeV, and 8.482 MeV, respectively. 

The third process is
\begin{equation}
\bar{\nu}_{e}+^{9}{\rm Be} \rightarrow \bar{\nu}_{e}+^{5}{\rm Li}+^{4}{\rm He} \rightarrow \bar{\nu}_{e}+2\times ^{4}{\rm He}+{\rm H}. 
\label{16}
\end{equation}
The binding energies of the nuclei $^{9}{\rm Be}$ and $^{5}{\rm Li}$ are 58.164 MeV and 27.633 MeV, respectively. 

The fourth process is
\begin{equation}
\bar{\nu}_{e}+^{19}{\rm F} \rightarrow \bar{\nu}_{e}+^{15}{\rm N}+^{4}{\rm He}. 
\label{17}
\end{equation}
The binding energies of the nuclei $^{19}{\rm F}$ and $^{15}{\rm N}$ are 147.801 MeV and 115.492 MeV, respectively.

The threshold energies of nucleus splitting by reactor antineutrinos, the weighted integral cross sections and the contributions to the nucleus splitting rate from electroweak and semi-weak interactions are presented in table 6.

\begin{center}
{{\bf Table 6.} The contributions from the semi-weak and electroweak interactions }\\
\begin{tabular}{cccccc}
\multicolumn{6}{c}{to the rates of nucleus splitting by reactor antineutrinos} \\ 
\hline
\multicolumn{1}{c}{Nucleus Y}
&\multicolumn{1}{c}{$E_{\rm th}^{\rm Y}$}
&\multicolumn{1}{c}{$\Sigma^{\rm Y}_{\rm Z}$}
&\multicolumn{1}{c}{$\Sigma^{\rm Y}_{\rm ps}$} 
&\multicolumn{1}{c}{${\cal N}^{\rm Y}_{\rm Z}$} 
&\multicolumn{1}{c}{${\cal N}^{\rm Y}_{\rm ps}$} \\
& (MeV) & (cm$^{2}$/fission) & (cm$^{2}$/fission) & 1/(mole$\cdot$day) & 1/(mole$\cdot$day) \\
\hline
$^{3}$He & 5.493 & $1.18 \cdot 10^{-45}$ & $1.08 \cdot 10^{-39}$ & $7.4 \cdot 10^{-5}$ & 68 \\
$^{7}$Li & 2.467 & $3.64 \cdot 10^{-43}$ & $3.98 \cdot 10^{-38}$ & 0.023 & $2.5\cdot 10^{3}$ \\
$^{9}$Be & 2.235 & $7.38 \cdot 10^{-43}$ & $4.97 \cdot 10^{-38}$ & 0.046 & $3.1\cdot 10^{3}$ \\
$^{19}$F & 4.014 & $3.24 \cdot 10^{-43}$ & $7.41 \cdot 10^{-39}$ & 0.020 & $4.7\cdot 10^{2}$ 
\\
\hline
\end{tabular}
\end{center}

An answer to the question on how much realistic the observation of the processes (\ref{14})--(\ref{17}) is at the nucleus splitting rates given in table 6 could probably be expected from chemists together with experts in atomic physics. 

\begin{center}
{\large \bf 6. Conclusion}
\end{center}

In my opinion, an experimental study of whether the semi-weak interaction influences the fission rates of long-lived transuranic isotopes irradiated by reactor antineutrinos would also be of interest.

A question remains completely open on whether the muon neutrinos possess semi-weak interaction and, if so, what is their Yukawa coupling constant with the pseudoscalar boson $\varphi_{ps}$.

I am sincerely grateful to S.P. Baranov for the numerous discussions of problems connected with the present work.

\end{small}
\end{document}